\documentclass[reprint,showpacs,amsmath,amssymb,superscriptaddress,nofootinbib,aps,twocolumn]{revtex4}
\usepackage{multirow}
\usepackage{bigstrut}
\usepackage{amssymb}
\usepackage{amsmath}
\usepackage{graphicx}
\usepackage{dcolumn}
\usepackage{bm}
\usepackage{relsize}
\usepackage{cancel}
\usepackage{color}
\usepackage{float}

\begin{document}

\title{Fast control of semiconductor qubits beyond the rotating-wave approximation}

\author{Yang Song}
\affiliation{Condensed Matter Theory Center, Department of Physics, University of Maryland, College Park, Maryland 20742, USA}
\author{J.~P.~Kestner}
\affiliation{Department of Physics, University of Maryland Baltimore County, Baltimore, Maryland 21250, USA}
\author{Xin Wang}
\affiliation{Department of Physics and Materials Science, City University of Hong Kong, Tat Chee Avenue, Kowloon, Hong Kong SAR, China}
\author{S.~Das Sarma}
\affiliation{Condensed Matter Theory Center, Department of Physics, University of Maryland, College Park, MD 20742, USA}

\begin{abstract}
We present a theoretical study of single-qubit operations by oscillatory fields on various semiconductor platforms. We explicitly show how to perform faster gate operations by going beyond the universally-used rotating wave approximation (RWA) regime, while using only two sinusoidal pulses.  We first show for specific published experiments how much error is currently incurred by implementing pulses designed using standard RWA. We then show that an even modest increase in gate speed would cause problems in using RWA for gate design in the singlet-triplet (ST) and resonant-exchange (RX) qubits. We discuss the extent to which analytically keeping higher orders in the perturbation theory would address the problem. More strikingly, we give a new prescription for gating with strong coupling far beyond the RWA regime. We perform numerical calculations for the phases and the durations of two consecutive pulses to realize the key Hadamard and $\frac{\pi}{8}$ gates with coupling strengths up to several times the qubit splitting. Working in this manifestly non-RWA regime, the gate operation speeds up by two to three orders of magnitude and nears the quantum speed limit without requiring complicated pulse shaping or optimal control sequences.
\end{abstract}

\maketitle

\section{Introduction}

Recent experiments in spin qubit systems demonstrated full single-qubit control via electrical ac driving at microwave frequencies \cite{Medford_PRL13,Shulman_NatCom14, Laucht_SciAdv15, Veldhorst_Nat15, Kim_NPJ15}. Advantages of such an approach include the ability to operate the qubit while staying near a sweet spot where the charge noise is considerably suppressed \cite{Medford_PRL13, Kim_NPJ15} while simultaneously filtering out low-frequency noise off-resonance with the driving field.  However, designing an ac control pulse is not necessarily a trivial matter. There are a variety of approaches for designing pulses with some set of desirable qualities such as speed, finite bandwidth, low fluence, robustness against pulse length errors, etc.

The standard approach in spin qubit experiments is simply to use the rotating wave approximation (RWA). RWA is rather ubiquitous in a variety  of contexts, arising, for example, when a laser is used to couple two atomic states. When the driving field is near resonance with the transition frequency between the states, RWA gives a straightforward prescription for full control of the two-level system by manipulating the phase and the duration of the ac control pulse.  With the exception of extremely intense lasers, the amplitude of the ac control field is typically  small when compared to the resonant frequency, and is extensively employed in atomic and molecular physics.  However, semiconductor qubits have a much smaller two-level splitting than typical atomic systems and can be operated in a regime where RWA is no longer valid,  i.e., the driving ac field strength is not necessarily much smaller than the qubit energy spacing. In practice, semiconductor qubits are often operated with weak driving in order to take advantage of the simplicity of the RWA control design.  However, there is no reason stronger driving could not be used in conjunction with state-of-the-art pulse-shaping or some other control method to perform qubit operations. In fact, strong driving is attractive for semiconductor qubit operations because it leads to both faster gate operations and less exposure to background noise.  The latter is particularly important at this stage, as large-scale quantum computation requires errors per gate to be below a given threshold for fault-tolerant error correction (FTEC) \cite{Knill_PRSA98, Aharonov_SIAM08,Gottesman_thesis97}, so one needs gate times substantially shorter than the $T_2$ coherence time. This threshold is often estimated to be $\sim 10^{-4}$ \cite{Preskill_PRSA98, DiVincenzo_Fortschritte2000, Steane_PRA03, Knill_Nat10, Brown_PRA11}, and even for more recent surface codes with thresholds of $\sim 0.7\%$ \cite{Fowler_PRA12} the massive overhead is greatly reduced if one can reduce errors down to near the $10^{-4}$ level \cite{Martinis_15}. This fact also provides motivation to design gates accurately -- even in the absence of any environmental noise, using control pulses based on an RWA analysis in a strong driving regime would obviously introduce completely avoidable deterministic errors (which would have to be corrected introducing avoidable overhead in the quantum computing operations)! The objective of the first part of this work is to quantify the error due to a RWA pulse design by considering realistic experimental situations involving different qubit architectures. Such a quantification in the specific context of spin qubits is very important as a practical guide, and here we provide the necessary analysis.

\begin{table*}
\caption{\label{tab:exper_parameters}
Estimated infidelity due to RWA for different semiconductor qubit experiments. Single-spin qubit parameters are listed for donor electrons (e),  nuclear spins (n), and gate-confined quantum dots in silicon (Si).
}
\renewcommand{\arraystretch}{1.3}
\tabcolsep=0.9 cm
\begin{tabular}{ccccc}%
 \hline\hline
            &   $\omega_z/2\pi$   &  $\omega_x/2\pi$ & $1-\bar{\mathcal{F}}^{\rm Max}_{\rm rwa}$ & Reference \\
               \hline
RX qubit & 210-370MHz & 41-112MHz & 1.0-3.8$\times 10^{-3}$ & Ref.~\cite{Medford_PRL13}
\\
ST qubit & 60MHz & 20MHz & $4.6\times 10^{-3}$& Ref.~\cite{Shulman_NatCom14}
\\
Hybrid      & 11.5GHz & 110MHz & $3.8\times 10^{-6}$&  Ref.~\cite{Kim_NPJ15}
\\
\multirow{3}{*}{\renewcommand{\arraystretch}{1.0}$\begin{array}{c}\textrm{single}\\ \textrm{spin}\\ \textrm{qubit}\end{array}$}
                          &(e) 43GHz $\quad$ & 48kHz&  $5.2\times 10^{-14}$  &   \multirow{2}{*}{ Ref.~\cite{Laucht_SciAdv15}} \\
                             &(n) 97MHz $\quad$ & 10kHz& $4.6\times 10^{-10}$\\
                            &(Si)   39GHz & 730kHz&   $1.5\times 10^{-11}$ & Ref.~\cite{Veldhorst_Nat15}
\\
\hline\hline
\end{tabular}
\end{table*}

It turns out that to achieve faster gating for certain oscillatory driven quantum-dot systems such as the RX \cite{Medford_PRL13} or ST qubit \cite{Shulman_NatCom14}, one can no longer use RWA as a guide for constructing the control pulse. In the second part of this paper we carry out a numerical analysis of a specific form of control pulse and present the key pulse control parameters for experiments to achieve certain important gates with much higher speed.
We do not search for the shortest possible solution, as in optimal control theory, but instead we catalog all realizations within a simple, highly constrained space. Some elegant optimal control approaches apply to two \cite{Khaneja_PRA01} and three qubits \cite{Khaneja_PRA02}, but we consider only single-qubit operations. Other approaches focus on accounting for the RWA resonance shift \cite{Ashhab_PRA07, Lu_PRA12, Yan_PRA15, Romhanyi_PRB15}, but we wish to keep the entire dynamics, including the rapid counter-oscillating terms. There are a wealth of pulse-shaping approaches, such as analytical reverse-solving \cite{Barnes_PRL12, Barnes_SciRep15}, multi-parameter optimization schemes such as CRAB \cite{Doria_PRL11, Scheuer_NJP14}, and other iterative updating algorithms for pulse sequences such as GRAPE \cite{Khaneja_JMR05} and Krotov's approach \cite{Krotov_book95, Maximov_JCP08} based on optimal control theory. In comparison, our approach focuses on simple sinusoidal pulses with four parameters (the durations and phases of the two pulses), and our theoretical results will serve to guide a straightforward experimental calibration of the control pulses. Optimal control theory or pulse-shaping techniques may well be more desirable for a variety of reasons in a given case.  What we show in the second part of our work, though, is that one can already consider driving in the strongly non-RWA regime with a relatively simple, two-pulse, numerical construction. We believe that the specificity and the concreteness of our results as well as the simplicity of the proposed pulses make our work of immediate usefulness in the context of faster gate control in various semiconductor qubit experiments going on worldwide.

An overview of this paper is as follows. The first part of this paper quantifies the deterministic error induced using RWA pulse design in ST, RX, as well as spin-charge hybrid \cite{Kim_NPJ15} and different single-spin qubit \cite{Laucht_SciAdv15, Veldhorst_Nat15} systems.  Both the nominally resonant (when field frequency equals the two-level splitting) and nonzero detuning cases are covered.  We find that for the RX qubit \cite{Medford_PRL13, Taylor_PRL13} and the resonantly-driven ST qubit \cite{Shulman_NatCom14,Levy_PRL02, Klauser_PRB06}, the deterministic RWA-induced errors are already larger than the fault tolerance threshold with the experimentally reported parameters, which grow even larger if the gates are implemented faster in an attempt to reduce random environment-induced errors. Therefore special attention will be paid to these two cases. However, the analysis applies generally to any strongly driven two-level system, and should be applicable to other qubit systems where fast gate operations are desirable.

Following the numerical analysis, we apply higher-order perturbation theory (two orders beyond the leading order, which is RWA) on the converted Floquet Hamiltonian \cite{Shirley_thesis63,Shirley_PRB65} and obtain an analytical expression for the state-averaged gate infidelity induced by using only an RWA (i.e. the zeroth-order) analysis to design gate pulses. We show that there is an intermediate parameter regime where incorporating higher-order corrections in the gate suppresses the deterministic error by orders of magnitude. This is quite relevant to the currently on-going ST and RX experiments, and could be useful for pulse designs to also reduce random errors, as previously done with piecewise constant pulses \cite{Wang_NatCom12,  Kestner_PRL13}.

In the second part of this work, we go beyond perturbative approaches such as RWA, and solve numerically exactly for pulse control parameters that realize important single-qubit gates for a large range of driving strengths. This both eliminates deterministic error and increases gate speed, thus indirectly also reducing random error. Our solutions increase the gate speed by more than two orders of magnitude, making our results important for future experiments involving quantum error corrections in semiconductor qubits. We use a simple pulse sequence with only four parameters.  Experimental implementation in the nonideal solid-state environment would proceed by calibrating the pulses in a small region of parameter space near our numerical solution.

The structure for the rest of the paper is as follows (and we also provide a guidance for the key results). Section~\ref{subsec:RWA_sinu_numer} gives a systematic numerical quantification of the averaged RWA infidelity, with the underlying leading-order effect worked out analytically [Eq.~(\ref{eq:infidelity_rwa_ext})] in Sec.~\ref{subsec:analytic_extension}. The discussion is tailored to various key semiconductor qubit systems summarized in Table~\ref{tab:exper_parameters}, with each system's position in parameter space clearly mapped out in Fig.~\ref{fig:4}.  Section~\ref{sec:fast_gate} works out the exact pulse control parameters to realize Hadamard and $\frac{\pi}{8}$ gates beyond the RWA regime, with key results summarized in Figs.~\ref{fig:Sinu_Hada} and \ref{fig:Sinu_PiOver8}. An extension to shifted oscillatory coupling is demonstrated in Appendix~\ref{app:shifted_sinu}. Section~\ref{sec:conclusion} concludes this work, where a comparison of our gate speed to the quantum speed limit is also given.

\section{Analysis of RWA errors in semiconductor qubit systems}\label{sec:Error_analysis}

We start with the Hamiltonian for a quantum two-level system,
\begin{eqnarray}
H_{\rm sin}= \frac{J_z}{2}\sigma_z + \frac{J_x(t)}{2} \sigma_x, \label{eq:H_exact}
\end{eqnarray}
where the bases of the Pauli matrices $\sigma_{x,z}$ are the qubit states $|0\rangle$ and $|1\rangle$, $J_z= \hbar \omega_z$ is the two-level splitting in the absence of driving, and the subscript ``$\rm sin$'' indicates that the off-diagonal control field $J_x(t)$ has the sinusoidal form
\begin{eqnarray}\label{eq:J_x_t}
J_x(t)= \hbar\omega_x \cos (\omega t+\Phi).
\end{eqnarray}
with a driving angular frequency $\omega$ and phase $\Phi$. When $\omega$ is tuned close enough to $\omega_z$, and the coupling strength is weak ($|J_x/ J_z|\ll 1$), one may employ the well-known rotating wave approximation by ignoring the so-called counter-rotating part of $J_x$ in Eq.~(\ref{eq:J_x_t}). The approximated Hamiltonian is therefore \cite{Rabi_PR37, Bloch_PR40, Vandersypen_PMP04}
\begin{eqnarray}\label{eq:H_rwa}
H_{\rm rwa} &=& \frac{J_z}{2}\sigma_z + \frac{\hbar\omega_x}{4} \left[ e^{-i(\omega t+\Phi)}\frac{\sigma_+}{2} +e^{i(\omega t +\Phi)}\frac{\sigma_-}{2}\right]
\nonumber\\
&=& \hbar \left[\frac{\omega_z}{2}\sigma_z + \frac{\omega_x}{4}  \hat{\mathbf{b}}(t)\cdot\bm\sigma \right]
\end{eqnarray}
in the same laboratory frame of Eq.~\eqref{eq:H_exact}, where the conventional notations $\sigma_\pm\equiv\sigma_x\pm i\sigma_y$, $\bm\sigma\equiv\sigma_x \hat{\mathbf{x}} + \sigma_y \hat{\mathbf{y}} + \sigma_z \hat{\mathbf{z}} $, and $\hat{\mathbf{b}}(t)=\cos(\omega t+\Phi)\hat{\mathbf{x}}+\sin(\omega t+\Phi)\hat{\mathbf{y}}$. Obviously, Eq.~(\ref{eq:H_rwa}) is much easier to use [than Eq.~(\ref{eq:H_exact})] in interpreting and understanding experiments, and therefore, in many applications, Eqs.~(\ref{eq:H_exact}) and (\ref{eq:H_rwa}) are considered equivalent (often uncritically), and this is referred to as RWA in the literature.

In this section we examine how good this approximation is for the range of parameters $\omega_z$, $\omega_x$ and $\omega$ actually used in various semiconductor qubit systems.  Note that the Rabi frequency becomes $\omega_x/2$ upon dropping the counter-rotating part of driving terms.  We therefore use the corresponding nominal Rabi period $T_R\equiv 4\pi/\omega_x$ to compare resonant and off-resonant cases where applicable.

We focus on the semiconductor qubit platforms where oscillatory control fields may be or have already been utilized. Most attention is given to the RX (Sec.~\ref{subsec:RWA_sinu_numer}) and various ST qubit systems (Sec.~\ref{subsec:RWA_sinu_numer} and more in Sec.~\ref{sec:fast_gate}), which can be operated in the strong coupling regime where the RWA may break down. In comparison, we also study other systems such as spin-charge hybrid qubits and different kinds of single-spin qubits, for which we reconfirm that the RWA is safe for the control parameters concerned (even from the stringent quantum error correction constraint considerations). However, we will show quantitatively how efforts to speed up gate operations for scalable quantum computation ($T_R<\{T_2/10^3, T^*_2\}$, $T^*_2$ being the ensemble coherence time \cite{Hanson_RMP07}) would enlarge the error due to using RWA, and we give a detailed numerical study in Sec.~\ref{subsec:RWA_sinu_numer}. Focusing on the leading orders of the errors, we present analytical results that can be used to greatly reduce errors as one leaves the strong coupling regime in Sec.~\ref{subsec:analytic_extension}.

\subsection{Systematic numerical study of RWA infidelity}\label{subsec:RWA_sinu_numer}

As an $SU(2)$ unitary matrix, the evolution operators corresponding to the Hamiltonian in Eqs.~\eqref{eq:H_exact} and \eqref{eq:H_rwa} may be written as
\begin{eqnarray}\label{eq:U_axial_way}
U_\lambda(t) &=& \exp\left[-i \hat{\mathbf{n}}_\lambda(t)\cdot \bm{\sigma} \frac{\phi_\lambda(t)}{2}\right]
\nonumber\\
&=& \cos\frac{\phi_\lambda(t)}{2} -i \hat{\mathbf{n}}_\lambda(t) \cdot \bm{\sigma} \sin\frac{\phi_\lambda(t)}{2},
\end{eqnarray}
where $\hat{\mathbf{n}}$ and $\phi$ can be physically interpreted as the rotation axis and angle, respectively, and $\lambda$ is used to denote ``$\rm sin$'' or ``$\rm rwa$''. The deviation due to RWA from the desired rotation can then be expressed as
\begin{eqnarray}\label{eq:delta_U_rwa}
\delta U_{\rm rwa} = U^\dag_{\rm rwa} U_{\rm sin} =\exp\left[-i \delta\hat{\mathbf{n}}\cdot \bm{\sigma} \frac{\delta\phi}{2}\right],
\end{eqnarray}
where
\begin{eqnarray}
\cos\frac{\delta\phi(t)}{2} &=& \cos\frac{\phi_{\rm rwa}}{2} \cos\frac{\phi_{\rm sin}}{2} \nonumber\\
 &&+ \hat{\mathbf{n}}_{\rm rwa}\cdot \hat{\mathbf{n}} _{\rm sin} \sin\frac{\phi_{\rm rwa}}{2} \sin\frac{\phi_{\rm sin}}{2},
\end{eqnarray}
and a similar relation follows for $\delta\mathbf{n}(t)$ \cite{footnote_delta_n}.
In this work we evaluate this error using the state-averaged infidelity $1-\bar{\mathcal{F}}_{\rm rwa}$, straightforwardly defined as \cite{Bruss_PRA98,Bowdrey_PLA02}
\begin{eqnarray}\label{eq:benchmark_2}
1\!-\!\bar{\mathcal{F}}_{\rm rwa}\!\!&=&\! 1\!-\! \frac{1}{4\pi} \int \!d\Omega {\rm Tr}\left[\left(\rho_\Omega \delta U_{\rm rwa}\right)\left(\rho_\Omega \delta U_{\rm rwa}^\dag\right) \right]
\nonumber\\
&=&\frac{1}{2}\!-\!\frac{1}{3}\!\! \sum_{i=x,y,z}\! {\rm Tr}\left[\!\left(\frac{\sigma_i}{2}\delta U_{\rm rwa}\right) \!\left(\frac{\sigma_i}{2} \delta U_{\rm rwa}^\dag\right) \right]
\end{eqnarray}
where $\rho_\Omega$ is the density matrix for a spinor with the solid angle $\Omega$. Using Eq.~(\ref{eq:delta_U_rwa}) the above equation can be expressed as
\begin{eqnarray}\label{eq:infidelity_in_delta_phi}
1-\bar{\mathcal{F}}_{\rm rwa}= \frac{2}{3} \sin^2\frac{\delta\phi(t)}{2}.
\end{eqnarray}
It is clear that the maximum infidelity is 2/3 from Eq.~(\ref{eq:infidelity_in_delta_phi}). In the following, we study individually different qubit systems for the error induced by using RWA to design the operation.

\begin{figure}[!htbp]
\centering
\includegraphics[width=8.5cm]{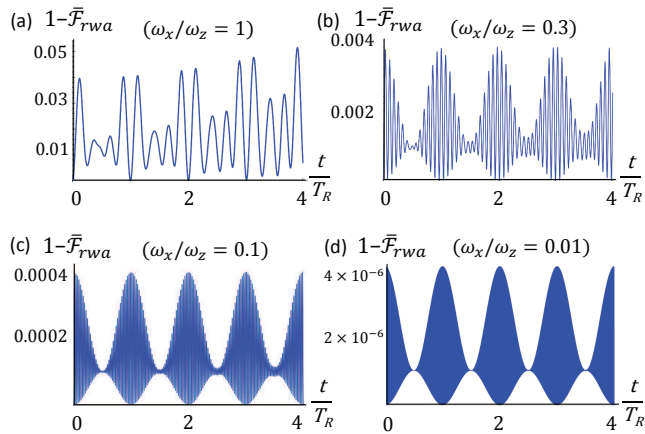}
\caption{Averaged infidelity due to RWA, $1-\bar{\mathcal{F}}_{\rm rwa}$,  as a function of time  for the zero detuning case ($\omega=\omega_z$). Results for various driving strengths $\omega_x/\omega_z$ are shown: (a) 1, (b) 0.3, (c) 0.1, and (d) 0.01.} \label{fig:1}
\end{figure}

\textit{Resonant exchange (RX) qubit.} The resonant-driven version of the exchange-only triple-dot qubits \cite{DiVincenzo_Nat00, Laird_PRB10, Gaudreau_NatPhys12, Medford_NatNano13, Braakman_NatNano13} has been developed \cite{Medford_PRL13, Taylor_PRL13} with the intention to filter out low-frequency charge noises. In the experiment of Ref.~\cite{Medford_PRL13},  the control fields can be expressed phenomenologically as $J_z \approx p^2/\epsilon_0$ and $J_x\approx \sqrt{3} p^2 \epsilon/\epsilon_0^2$ where $p$ is the tunnel coupling between the neighboring quantum dots, $\epsilon_0$ measures the voltage shift of the middle gate, and the oscillatory $\epsilon(t)$ is the relative detuning with respect to the center of the (111) charge region. In the RWA regime  the key quantity $\omega_x/\omega_z=J_x/J_z\approx \sqrt{3} \epsilon/\epsilon_0$. From the parameters provided in Ref.~\cite{Medford_PRL13} we have found that $\omega_x/\omega_z$ is around 15\%-30\% (see Table~\ref{tab:exper_parameters} for more details), corresponding to an RWA error of  approximately $0.4\%$ [cf. Fig.~\ref{fig:1}(b)]. While this error is small compared to those caused by nuclear and charge noises, it still exceeds the typical target threshold of $10^{-4}$ for quantum error correction, meaning that even after the environmental noises have been eliminated (e.g. through dynamical decoupling or other techniques), the RWA remains a hurdle to fault tolerant quantum computing. Unlike environmental noise, the RWA error is systematic and must be treated in a different way than standard techniques such as dynamical decoupling which are typically used to mitigate environmental decoherence. This is precisely what motivates our work.

We further discuss the ranges of the coupling strengths in the RX qubit and their consequences on RWA. Reference~\cite{Medford_PRL13} has demonstrated that by increasing the middle gate voltage, both the qubit splitting $J_z$ and the resonant coupling $J_x$ increase. The latter however increases much faster: While $\omega_z/2\pi$ is increased from 0.355 to 1.98 GHz (a factor of $\sim5$), $d(\omega_x/4\pi) /dV_l$ is increased from 0.07 GHz/mV to 5  GHz/mV (a factor of $\sim70$). This indicates that while enhancing the control fields is appealing due to potentially faster gate operation and smaller influence from noise, it will actually lead to even larger RWA error. The corresponding infidelity is a few percent [cf. Fig.~\ref{fig:1}(a)] or larger, which can be comparable to those contributed from environmental noises.

On the other hand, if $\omega_x$ is kept small, the gate speed becomes an issue. Typically one requires that each quantum gate operation is at least $10^3$ times faster then the coherence time, which sets a lower bound on $\omega_x$. This is estimated below. When $\omega_z/2\pi=0.2$ GHz, quantum coherence time $T_2$ reaches 20 $\mu$s as measured experimentally \cite{Medford_PRL13}, which imposes $\omega_x/4\pi \geq 10^3/T_2 = 0.05$ GHz. This  ratio $\omega_x/\omega_z = 1/2$ gives a fairly good estimate on how slow the Rabi oscillation is allowed to go for quantum computation, corresponding to an RWA infidelity of about 1\% which is unacceptably high. Ways to correct the problem include prolonging the coherence time and enlarging $\omega_z$ while keeping $\omega_x$ small by fine tuning the gate voltages, although other noises may play a role in the latter approach.

\textit{Singlet-Triplet (ST) qubit.} The ST qubit \cite{Petta_Science05} can similarly benefit from the oscillatory driving to filter out the quasi-static charge noise. In order to conform with the Hamiltonian Eq.~\eqref{eq:H_exact}, we take $|\uparrow\downarrow\rangle$ and $|\downarrow\uparrow\rangle$ as the qubit bases so that  $J_z=g^* \mu_B \Delta B_z$ corresponds to the magnetic field gradient and $J_x$ is the exchange interaction between the singlet and triplet states which can be made oscillatory. Such oscillating control has been demonstrated in Ref.~\cite{Shulman_NatCom14}, although the experiments are done for the different purpose of dynamical Hamiltonian estimation\cite{Klauser_PRB06}. In this experiment, $\omega_x/\omega_z$ is estimated to be approximately $1/3$ which means that the RWA error is large [cf. Table~\ref{tab:exper_parameters} and Fig.~\ref{fig:1}(b)]. One contributing factor to the large ratio of $\omega_x/\omega_z$ is the small $\Delta B_z$ one is able to maintain in this experiment. This factor has since improved to reach about a few GHz for $\omega_z$\cite{Yacoby_footnote} meaning that the RWA error can be greatly reduced.

The ST qubits may also be implemented across different platforms including GaAs and Si. While many of these qubits are being operated using piecewise constant pulses  \cite{Foletti_NatPhys09, Petersen_PRL13, Wu_PNAS14}, migration to oscillating pulses is straightforward thanks to the efficient electrical control over $J_x$ using gate voltages. In order to gain insight into the RWA error if these qubits were to be operated using oscillating pulses, we summarize the range of parameters $\omega_x$ and $\omega_z$ in Table~\ref{tab:parameter_ST}.  We have included cases with different methods to generate the magnetic field gradient, including dynamical nuclear spin polarization\cite{Foletti_NatPhys09,Shulman_NatCom14}, micromagnets in GaAs\cite{Petersen_PRL13},  micromagnets in Si \cite{Wu_PNAS14}, and the gradient produced by the donor hyperfine fields\cite{Kalra_PRX14}. From the table we can see that one has considerable freedom to tune the ratio $\omega_x/\omega_z$ in a wide range. In order to keep the RWA error below $10^{-3}$ we must have  $\omega_x/\omega_z \lesssim 0.1$, which however implies slow gates. In Sec.~\ref{sec:fast_gate}, we shall present a method to avoid using RWA, allowing access to large ratios of $\omega_x/\omega_z$ without introducing errors and implementing precisely-designed fast gates.

\begin{table}
\caption{\label{tab:parameter_ST} Representative parameters $\omega_z/2\pi=g^* \mu_B \Delta B_z/2\pi\hbar$ and the gate controlled exchange coupling amplitude $\omega_x/2\pi$  from the major recent ST systems with static control field. The shortcut  notations are, (D)NP=(dynamical) nuclear polarization and  m.magnet=micromagnet.}
\renewcommand{\arraystretch}{2.0}
\tabcolsep=0.15 cm
\begin{tabular}{cccc}%
 \hline\hline
            &   $\omega_z/2\pi$   &  $\omega_x/2\pi$  & Reference \\
               \hline
DNP in GaAs  & 1 GHz & $\leq 25$ \!GHz &  Ref.~\cite{Foletti_NatPhys09}
\\
\renewcommand{\arraystretch}{1.0}$\begin{array}{c}\textrm{NP by m.magnet}\\ \textrm{ in GaAs}\end{array}$
 & 3 GHz & $\leq 25$ \!GHz &  Ref.~\cite{Petersen_PRL13}
\\
m.magnet in Si/SiGe   & 0.014 GHz & $\leq 25$ \!GHz &  Ref.~\cite{Wu_PNAS14}
\\
\renewcommand{\arraystretch}{1.0}$\begin{array}{c}\textrm{P donor hyperfine}\\ \textrm{in Si(proposal)}\end{array}$   & 0.12 GHz & $\leq 1.2$ \!GHz &  Ref.~\cite{Kalra_PRX14}
\\
\hline\hline
\end{tabular}
\end{table}

\begin{figure}[!htbp]
\centering
\includegraphics[width=8.5cm]{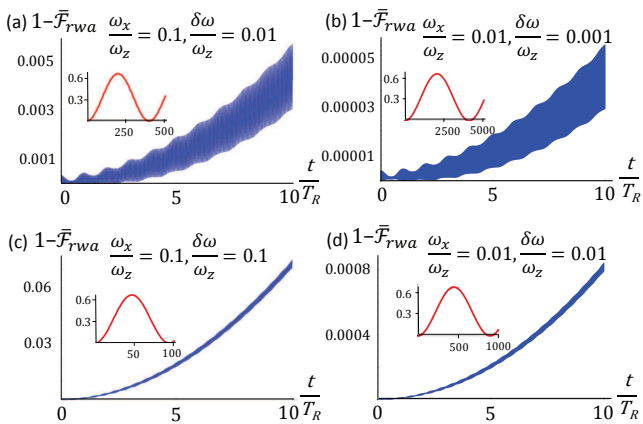}
\caption{Averaged infidelity due to RWA, $1-\bar{\mathcal{F}}_{\rm rwa}$,  as a function of time, for different relative detunings ($\delta \omega/\omega_z$ where $\delta \omega=\omega-\omega_z$) and coupling strengths ($\omega_x/\omega_z$), which are, respectively, (a) 0.1 and 0.01, (b) 0.01 and 0.001, (c) 0.1 and 0.1, and (d) 0.01 and 0.01. The insets plot $1-\bar{\mathcal{F}}_{\rm rwa}$'s behavior in the long time limit where another time period emerges.} \label{fig:2}
\end{figure}

\textit{Spin-charge hybrid qubit.} Recently the microwave-driven gate operation \cite{Kim_NPJ15} of the hybrid qubit \cite{Kim_Nat14} has been experimentally realized in Si nanostructures. As in the RX qubit, ac control of the hybrid qubit provides straightforward two-axis manipulation while suppressing the charge noise. In the experiment of Ref.~\cite{Kim_NPJ15} a $55\times 2\pi$ MHz Rabi oscillation is performed with the resonant frequency of 11.52 GHz. This $\omega_x/\omega_z$ ratio coincides with Fig.~\ref{fig:1}(d) which shows a negligible RWA error ($1-\bar{\mathcal{F}}_{\rm rwa}<5\times 10^{-6}$). However, this Rabi frequency is still relatively slow given the short $T_2$ time of 150 ns \cite{Kim_NPJ15}. To push for faster operations, possibly by dynamically modulating tunnel coupling \cite{Kim_NPJ15}, we expect that the RWA error increases to the case shown in Fig.~\ref{fig:1}(b). Again, the method we shall present in Sec.~\ref{subsec:analytic_extension} and Sec.~\ref{sec:fast_gate} will shed light on how one may reconcile the need of fast gate operation and low RWA error.

\textit{Single-spin qubit}. Single spin qubit manipulation utilizing  Kane's proposal \cite{Kane_Nat98} has been  realized in silicon with phosphorus ($^{31}\!$P) donors \cite{Laucht_SciAdv15}. Proximity gate voltage is demonstrated to modify the resonance frequencies of both the electron and nuclear spin under a static magnetic field $B_0$ via the Stark shift\cite{Laucht_SciAdv15}, as the change of localized electron distribution affects the effective $g$ factor and the hyperfine coupling. In this experiment, $B_0$ and the effective field from hyperfine interaction are equal or greater than $10^4$ of the oscillatory field amplitude $B_1$, with effective $\omega_z$ and $\omega_x$ shown in Table~\ref{tab:exper_parameters}. In this regime, RWA is clearly safe to use due to its negligible error. Moreover, the demonstrated Rabi frequency is already close to $2\pi\times 10^3/T_2$ with $T_2>10 (100)$ ms for electron (nuclear) spins \cite{Laucht_SciAdv15}, which have met the requirement of FTEC.

In another recent work using resonant-driven qubit control in non-purified silicon\cite{Veldhorst_Nat15}, a similar static field $B_0=1.4$ T has been used and the Rabi period of about 2.4 $\mu$s have been demonstrated. The $T_2$ time can reach $28$ ms, and together with a relatively stronger $J_z$ one may satisfy the FTEC requirement while keeping the RWA error negligible.

Having discussed several experimental platforms with the assumption that the driving frequency is on resonance with the qubit level splitting, we turn our attention to the detuning of driving frequency from exact resonance ($\delta \omega=\omega-\omega_z$), which is also an important RWA parameter and could affect the error. $\delta \omega$ is relevant in various qubit operations and calibrations.  Ramsey fringes \cite{Ramsey_PR50} have been used in several of the aforementioned experiments \cite{Shulman_NatCom14, Kim_NPJ15, Laucht_SciAdv15, Veldhorst_Nat15} to obtain ensemble coherence $T^*_2$ time with $\delta\omega$ reaching up to $0.1 \omega_z$ \cite{Shulman_NatCom14}, and to calibrate the resonance frequency $\omega_z$ \cite{Ramsey_PR50}.  Moreover, in the single-donor spin qubits \cite{Laucht_SciAdv15}, the detuning can be controlled by tuning $\omega_z$, as opposed to $\omega$, by proximity to the top gate. From Fig.~\ref{fig:2}, we see that when $\delta\omega/\omega_x$ reaches about 1/10, it starts to worsen the RWA appreciably. However, when $\delta \omega$ further increases such that $\delta\omega/\omega_x\gtrsim1$, its contribution to the infidelity stops growing: this is the far off-resonance regime where the qubit is not rotated. Another interesting feature is that the error due to detuning grows over time, which is absent in the zero-detuning case (cf. Fig.~\ref{fig:1}). This is relevant for the wide `non-interacting' region in the Ramsey experiment, as its mechanism assumes that the driving pulse keeps oscillating in the background to accumulate phase difference with Larmor precession in the RWA limit \cite{Ramsey_PR50}. We also observe that the shape of $1-\bar{\mathcal{F}}_{\rm rwa}$ versus $t/T_R$ is nearly the same when $\delta\omega/\omega_x$ is similar.  The insets in Fig.~\ref{fig:2} show the long time behavior of $1-\bar{\mathcal{F}}_{\rm rwa}$ where a much slower frequency emerges. There the upper bound of $1-\bar{\mathcal{F}}_{\rm rwa}=2/3$ [Eq.~(\ref{eq:infidelity_in_delta_phi})] is eventually reached.     These and other features can be explained by the leading-order analytical results we carry out, as presented in the next subsection.

\subsection{Error within a perturbative extension to RWA}\label{subsec:analytic_extension}

Analytical expressions of the qubit evolution are useful in quantum gate operations for many reasons. In particular, they are simple to use and avoid the numerical difficulty of integrating oscillating functions. Here we apply an extended higher-order analytical approximation \cite{Shirley_PRB65,Shirley_thesis63} and look deeper into the error caused by RWA.

We follow the theory developed by Shirley and others\cite{Shirley_PRB65,Shirley_thesis63} for going beyond RWA in a perturbative expansion. The periodic time-dependent Hamiltonian is transformed into a time-independent Floquet Hamiltonian, which is then solved by successive applications of quasi-degenerate perturbation theory in terms of the small parameters involved in this problem: $\omega_x/ \omega_z$  and $\delta\omega/\omega_z$. To do this, we start with an exact transformation
\begin{eqnarray}
\tilde{H}_{\rm sin}=U_\omega H_{\rm sin} U_\omega^\dag + i\hbar \frac{\partial U_\omega}{\partial t} U_\omega^\dag
\end{eqnarray}
where $U_\omega=e^{i\omega\sigma_z t/2}$ converts the reference frame to the rotating one. In this frame, the eigenstate becomes quasi-degenerate with $\omega\approx \omega_z$ in the absence of the driving term, and
\begin{eqnarray}\label{eq:H_sinu_rotating}
\tilde{H}_{\rm sin}= \frac{\hbar}{2}\left(\begin{array}{cc}
\omega_z-\omega & \frac{1}{2} \omega_x(1+ e^{2i \omega t}) \\
\frac{1}{2} \omega_x(1+ e^{-2i \omega t}) & -\omega_z+\omega
\end{array}\right).
\end{eqnarray}
Dropping the fast oscillating terms $e^{\pm 2i\omega t}$ reduces the Hamiltonian to the RWA one. The exact equation for the evolution operator in this frame should follow
\begin{eqnarray}\label{eq:exact_HU_rot}
i\hbar \frac{\partial \tilde{U}_{\rm sin}(t)}{\partial t} &=& \tilde{H}_{\rm sin} \tilde{U}_{\rm sin}(t).
\end{eqnarray}
Using the unitarity of $U$, $U_{22}=U^*_{11}$ and $U_{12}=-U^*_{21}$, where $U_{11}$ and $U_{12}$ are the Cayley-Klein parameters, the matrix equation Eq.~(\ref{eq:exact_HU_rot}) reduces to one for a vector state $[\tilde{U}_{\rm sinu,11}(t), \tilde{U}_{\rm sinu,21}(t)]^T$ with the initial conditions $U_{\rm sinu,11}(0)=1$ and $U_{\rm sinu,21}(0)=0$.

By transforming  Eq.~(\ref{eq:exact_HU_rot}) into a corresponding Floquet equation with basis states incorporating the $e^{2i n\omega t}$ factor \cite{Shirley_PRB65,Shirley_thesis63}, a second-order perturbation analysis gives two general solutions, $(\tilde{U}_{\rm ext,11}(t), \tilde{U}_{\rm ext,21}(t))^T$,
\begin{eqnarray}\label{eq:nu_1}
\nu_1\!=\!\left(\!\!\begin{array}{c}
\sqrt{\Delta\omega'-\delta \omega'}- \frac{\omega_x}{8\omega} \sqrt{\Delta\omega'+\delta \omega'} e^{2i \omega t}
\\
\sqrt{\Delta\omega'+\delta \omega'}+ \frac{\omega_x}{8\omega} \sqrt{\Delta\omega'-\delta \omega'} e^{-2i \omega t}
\end{array} \!\!\right) \!\!\!\frac{e^{-i\frac{\Delta \omega''t}{2}}}{\sqrt{2\Delta\omega'}}
\end{eqnarray}
and
\begin{eqnarray}\label{eq:nu_2}
\nu_2\!=\!\left(\!\!\begin{array}{c}
\sqrt{\Delta\omega'+\delta \omega'}+ \frac{\omega_x}{8\omega} \sqrt{\Delta\omega'-\delta \omega'} e^{2i \omega t}
\\
-\sqrt{\Delta\omega'-\delta \omega'}+ \frac{\omega_x}{8\omega} \sqrt{\Delta\omega'+\delta \omega'} e^{-2i \omega t}
\end{array} \!\!\!\!\right) \!\!\!\!\frac{e^{i\frac{\Delta \omega''t}{2}}}{\sqrt{2\Delta\omega'}}
\end{eqnarray}
where
\begin{eqnarray}
&\delta \omega' = \delta\omega \left(1-  \frac{\omega_x^2}{64 \omega^2} \right)-\frac{\omega_x^2}{16 \omega},&
\nonumber\\
&\Delta\omega' = \sqrt{\frac{\omega_x^2}{4}+\delta \omega'^2}, \quad \Delta \omega''= \left(1-\frac{\omega_x^2}{64\omega^2}\right) \Delta\omega' .&
\end{eqnarray}
The well known first-order correction to the physical resonance frequency, the Bloch-Siegert shift $\omega_x^2/16 \omega$ \cite{Bloch_PR40}, is the last term of $\delta \omega'$. Considering the initial conditions for $\tilde{U}_{\rm sin}$, we have
\begin{eqnarray}\label{eq:U_solution_2nd}
\left(\begin{array}{c}
\tilde{U}_{\rm ext,11}(t) \\ \tilde{U}_{\rm ext,21}(t) \end{array} \right)
= C_1 \nu_1+ C_2 \nu_2,
\end{eqnarray}
with constant coefficients
\begin{eqnarray}
C_1 \!\!&=&\!\! \left(1+ \frac{\omega_x^2}{64\omega^2}\right)^{-1} \!\!\frac{\sqrt{\Delta\omega'-\delta \omega'}- \frac{\omega_x}{8\omega} \sqrt{\Delta\omega'+\delta \omega'} }{\sqrt{2\Delta\omega'}},\qquad\label{eq:coef_C1}
\\
C_2\! \!&=& \!\!\left(1+ \frac{\omega_x^2}{64\omega^2}\right)^{-1} \!\! \frac{\sqrt{\Delta\omega'+\delta \omega'}+ \frac{\omega_x}{8\omega} \sqrt{\Delta\omega'-\delta \omega'} }{\sqrt{2\Delta\omega'}}.\qquad\label{eq:coef_C2}
\end{eqnarray}
Taking the $\{\delta\omega, \omega_x\}/\omega_z\rightarrow 0$ limit recovers the results from RWA. Apart from the higher-order correction to the resonance shift, we see from Eqs.~(\ref{eq:nu_1}) and (\ref{eq:nu_2}) that the oscillating terms $e^{\pm 2i \omega t}$ also play a major role, the contribution of which can be readily identified from the oscillating nature of curves shown in Figs.~\ref{fig:1} and \ref{fig:2}.

Converting the solution $U_{\rm ext}$ implied from Eq.~(\ref{eq:U_solution_2nd}) into the form of Eq.~(\ref{eq:U_axial_way}), we can express the infidelity corresponding to $U_{\rm ext}$ in terms of the deviation angle $\delta\phi$ as Eq.~\eqref{eq:infidelity_in_delta_phi}. In the RWA limit, $\tilde{\phi}_{\rm rwa}=\Delta\omega t$ and $\hat{\tilde{\mathbf{n}}}_{\rm rwa}=(\omega_x,0,-2\delta\omega)$ [$\Phi$ in Eq.~(\ref{eq:J_x_t}) has been set to be 0].

\begin{figure}[!htbp]
\centering
\includegraphics[width=8.5cm]{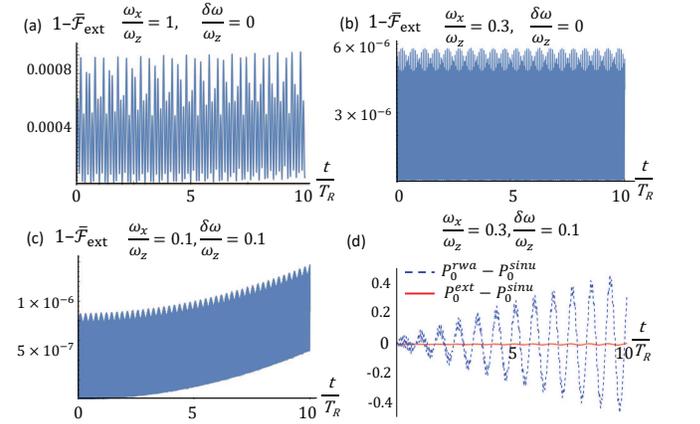}
\caption{Averaged infidelity $1-\bar{\mathcal{F}}_{\rm ext}$ from Eq.~(\ref{eq:U_solution_2nd}). The parameters in panels (a)-(c) are the same as in Figs.~\ref{fig:1}(a) and \ref{fig:1}(b) and Fig.~\ref{fig:2}(c), respectively. In (d) we plot the difference in the return probability between the exact solution and both the RWA and the extended analytical results for an initial $|0\rangle$ state and  $\omega_x/\omega_z=0.3$ and $\delta \omega/\omega_z=0.1$. } \label{fig:3}
\end{figure}

The infidelity of $U_{\rm ext}$ compared to the exact evolution $U_{\rm sin}$, $1-\bar{\mathcal{F}}_{\rm ext}$, is much smaller than that corresponding to RWA $1-\bar{\mathcal{F}}_{\rm rwa}$, as expected. This can also be seen from Fig.~\ref{fig:3}. The infidelities $1-\bar{\mathcal{F}}_{\rm ext}$ as shown in Figs.~\ref{fig:3}(a)  and \ref{fig:3}(c) are orders of magnitude smaller than what is shown in Figs.~\ref{fig:1} and \ref{fig:2}, which are well within the quantum error correction threshold. Even for very large $\omega_x/\omega_z$ ratio  [cf. Fig.~\ref{fig:3}(a)], the second-order perturbation results show a small error $\sim 8\times 10^{-4}$. In Fig.~\ref{fig:3}(d), we also present the improved accuracy using the return probability for the initial $|0\rangle$ state, $P_0$. Again we see a substantially smaller error for $U_{\rm ext}$ than for $U_{\rm rwa}$. Thus, the analytical perturbative results should be applicable in many practical situations already giving substantially more accurate results than RWA.

As the analytical solutions presented above satisfactorily approximate the exact sinusoidal result, we shall use them to investigate the  RWA error across a wider parameter range. Tracing the RWA error versus time, three different frequency components emerge and play important roles. As mentioned, a highly oscillatory component is always present with the counter-rotating frequency $2\omega$ in the rotating frame, and is the origin for the approximation in the RWA. This high frequency component is modulated by a combination of two lower frequencies, emergent from the interference of Floquet eigen-frequencies in the zeroth (RWA) and second-order perturbation. They correspond to $\Delta\omega''/2$ in the last factor of the two general solutions [Eqs.~(\ref{eq:nu_1}) and (\ref{eq:nu_2})] and its limiting value in RWA $\Delta\omega=\sqrt{\omega_x^2/4+\delta \omega^2}$. They produce a slow and a relatively fast component,
\begin{multline}
\Delta\omega'' - \Delta\omega
=- \frac{\omega^2_x}{16\Delta\omega \omega_z} \!\Bigg[ \delta \omega +\frac{ \delta\omega^2\!+\! \Delta\omega^2}{4 \omega_z}
  -\frac{\omega^4_x}{128 \omega_z \Delta\omega^2}\Bigg] \\
  +\mathcal{O} \left[\left\{\frac{\delta \omega}{\omega_z}, \frac{\omega_x}{\omega_z}\right\}^5\right], \label{eq:small_freq}
\end{multline}
\begin{equation} \label{eq:mid_freq}
\frac{1}{2}(\Delta\omega'' + \Delta\omega)
\approx  \Delta\omega.
\end{equation}
These two frequencies explain the envelope modulations in Figs.~\ref{fig:1} and \ref{fig:2} (the faster one) and in the inset of Fig.~\ref{fig:2} (the slower one). Taking $\delta \omega/\omega_z=0.1$ and $\omega_x/\omega_z=0.01$ for example, the leading contribution from Eq.~(\ref{eq:small_freq}) makes  $\sim \omega_x/800$, matching $T\sim 400 T_R$ in the inset of Fig.\ref{fig:2} (a). Contrasting $\tilde{U}_{\rm rwa}$ and $\tilde{U}_{\rm ext}$ in the form of Eq.~(\ref{eq:benchmark_2}), we are able to obtain all leading-order terms for $1-\bar{\mathcal{F}}_{\rm rwa}$ in parameters $\frac{\delta \omega}{\omega_z}$ and $\frac{\omega_x}{\omega_z}$ after some algebra,
\begin{widetext}
\begin{eqnarray}\label{eq:infidelity_rwa_ext}
1\!-\!\bar{\mathcal{F}}_{\rm rwa}(t)\!&\!=\!&\! \frac{\omega^2_x}{96\Delta\omega^4 \omega_z^2} \left[
\frac{3\omega_x^4 \!+\!24\omega_x^2 \delta\omega^2 \!+\!64\delta\omega^4 }{32}\!+ \!\frac{1}{16} \omega_x^2 \Delta \omega^2 \delta \omega^2\;t^2
\!-\! \frac{\omega_x^2 \Delta\omega^2}{4} (\cos 2\omega t\! +\!\delta \omega \; t\; \sin2\omega t)
\right.\nonumber\\
&&\left. \qquad\quad
+(\delta\omega^2 \!+\!\Delta\omega^2)\! \left(\frac{1}{8}\omega^2_x\cos\Delta\omega t\!  -\!\Delta\omega^2 \cos2\omega t \cos\Delta\omega t\! -\!\delta\omega \Delta \omega \sin2\omega t \sin \Delta\omega t \! \right)\!
\right]\!+\!\mathcal{O} \left[\left\{\frac{\delta \omega}{\omega_z},\! \frac{\omega_x}{\omega_z}\right\}^3\right].\qquad
\end{eqnarray}
\end{widetext}

\begin{figure}[!htbp]
\centering
\includegraphics[width=8.5cm]{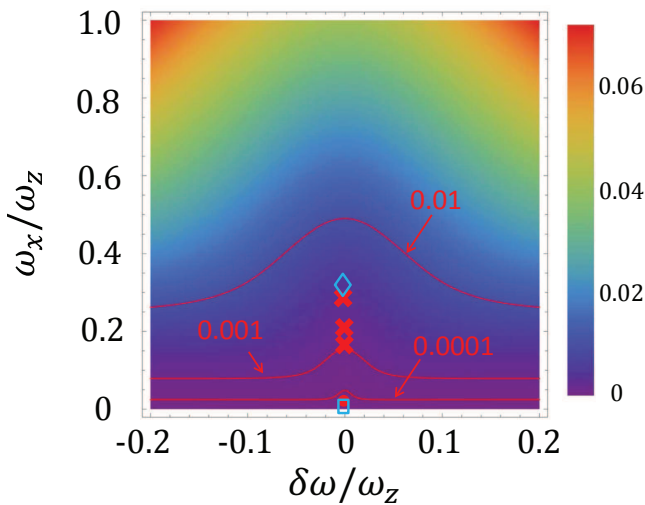}
\caption{Contour plot for the infidelity amplitude in the first Rabi cycle, $[1-\bar{\mathcal{F}}_{\rm rwa}]^{\rm Max}$ in Eq.~(\ref{eq:infidelity_Max}), as a function of relative detuning ($\delta \omega/\omega_z$) and  driving strength ($\omega_x/\omega_z$). We mark the $10^{-2}$, $10^{-3}$ and $10^{-4}$ contour lines. Three red crosses mark experimental operation points in the RX experiment (Figs. 2-4 in \cite{Medford_PRL13}), the light blue diamond marks the resonant ST experiment \cite{Shulman_NatCom14}, the red solid circle marks the hybrid qubit \cite{Kim_NPJ15}, and the light blue square marks the weakly driven single-spin experiments \cite{Laucht_SciAdv15,Veldhorst_Nat15} (see Table~\ref{tab:exper_parameters}).} \label{fig:4}
\end{figure}

From the equation above, we estimate the magnitude of error during one Rabi cycle as a function of $\frac{\delta \omega}{\omega_z}$ and $\frac{\omega_x}{\omega_z}$, which can be written as
\begin{widetext}
\begin{multline}\label{eq:infidelity_Max}
[1-\bar{\mathcal{F}}_{\rm rwa}]^{\rm Max} \approx  \frac{\omega^2_x}{96\Delta\omega^4 \omega_z^2} \left[
\frac{3\omega_x^4 +24\delta\omega^2 \omega_x^2 +64\delta\omega^4 }{32}
+ \frac{1}{16}\delta \omega^2 \omega_x^2 \Delta \omega T_R^2
+ \frac{\omega_x^2 \Delta\omega^2}{4} \sqrt{1 +(\delta \omega \; T_R)^2}\right.
\\\left. +(\delta\omega^2 +\Delta\omega^2)\sqrt{ \left(\frac{\omega^2_x}{8} +\Delta\omega^2\right)^2 +(\delta\omega \Delta \omega )^2}
\right].
\end{multline}
\end{widetext}
The numerical evaluation of Eq.~\eqref{eq:infidelity_Max} is summarized in Fig.~\ref{fig:4}, which gives an estimate of  the upper bound of the RWA  error for a wide range of parameters. Three lines are drawn where $[1-\bar{\mathcal{F}}_{\rm rwa}]^{\rm Max}$ coincides with the threshold values $10^{-4}$, $10^{-3}$ and $10^{-2}$. Several points in the parameter space relevant to experiments are also marked.

\section{Fast gate operations with arbitrary oscillatory driving amplitude}\label{sec:fast_gate}
\begin{figure}[h]
\centering
\includegraphics[width=8.5cm]{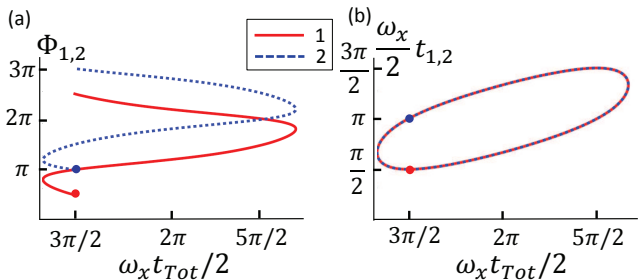}
\caption{Two-pulse solutions for the Hadamard gate in the resonant RWA limit. $\Phi_{1,2}$ and $t_{1,2}$ are the phases and durations of the two pulses. The solution is plotted parametrically, with (a) $\{\Phi_1,\Phi_2\}$, and (b) $\{t_1, t_2\}$ versus the total gate time, $t_{Tot}=t_1+t_2$. Times are scaled by the driving frequency, $\omega_x$.   Note that all four curves are closed or periodic. The usual RWA solution, $\{\Phi_1=\frac{\pi}{2},\Phi_2=\pi, \frac{\omega_x}{2}t_1=\frac{\pi}{2}, \frac{\omega_x}{2}t_2=\pi\}$ in our notation, is marked by solid dots.  These dots also give a sense of how to associate points on the multivalued curves at a given total time.
}\label{fig:gate_RWA}
\end{figure}

We now consider the construction of gate operations with strong driving outside the range of validity of the RWA (and its perturbative generalization). Two oscillatory pulses of the form $\omega_x \cos\left(\omega t_i + \Phi_i\right)$ are sufficient to achieve any single-qubit logic gate, which contains 3 degrees of freedom: two specifying the axis and one the rotation angle. Even though we have taken fixed driving frequency and amplitude, there are still 4 degrees of freedom in the two pulse sequence, i.e., the phase and duration of each segment.  We expect, then, that there is an infinite number of two-pulse realizations for a given logical gate, with the solution vector forming a curve in the 4D space $\{\Phi_1,\Phi_2, t_1,t_2\}$. We show below that this is indeed true, and select an optimum solution with the minimum total time.  This allows us to leave the weak coupling limit and increase the gate speed while still having simple pulse controls, as in RWA, as opposed to a complex pulse-shaping technique. This scheme should be useful for the current semiconductor qubit experiments.

The essential set of single-qubit gates necessary for universal quantum computation can be the Hadamard (H) gate and the `$\pi/8$' (T) gate, i.e., a $\pi$ rotation around $\hat{\mathbf{x}}+\hat{\mathbf{z}}$ direction and a $\pi/4$ rotation around $z$-axis, respectively. In the form compatible with $U_{22}=U^*_{11}$ and $U_{12}=-U^*_{21}$, one has
\begin{eqnarray}
U_{H} &=& \frac{1}{\sqrt{2}}\left(\begin{array} {cc} i& i \\ i & -i \end{array}\right), \label{eq:U_H}\\
U_{T} &=& \left(\begin{array} {cc} e^{-i\pi/8}& 0 \\ 0 & e^{i\pi/8} \end{array}\right),\label{eq:U_T}
\end{eqnarray}
up to an inconsequential global sign. Forming the H or T gate by two sinusoidal pulses amounts to searching for solutions to the differential Eq. \eqref{eq:exact_HU_rot} such that
\begin{eqnarray}\label{eq:U_2pulses}
\tilde{U}_{\rm sin}(t_2) \tilde{U}_{\rm sin}(t_1) = U_{H/T},
\end{eqnarray}
that is, to a root-searching problem,
\begin{eqnarray}
|\tilde{\mathbf{U}}_{I\!I}- \mathbf{U}_{H/T}|=0,
\end{eqnarray}
where the two vectors are
\begin{eqnarray}
\tilde{\mathbf{U}}_{\!I\!I}\!\!=\!\left( \!\begin{array}{c}
\!\tilde{U}_{\!{\rm sin},\!1\!1}(t_2) \tilde{U}_{\!{\rm sin},\!1\!1}(t_1\!) - \tilde{U}^*_{\!{\rm sin},\!2\!1}(t_2) \tilde{U}_{\!{\rm sin},\!2\!1}(t_1\!) \!\\
\!\tilde{U}_{\!{\rm sin},\!2\!1}(t_2) \tilde{U}_{\!{\rm sin},\!1\!1}(t_1\!) + \tilde{U}^*_{\!{\rm sin},\!1\!1}(t_2) \tilde{U}_{\!{\rm sin},\!2\!1}(t_1\!)\! \end{array}\! \right),
\end{eqnarray}
and $\mathbf{U}_{H(T)}=(U_{H(T),11},U_{H(T),21})^T$. Here the logical gates are realized in the rotating frame as they are in the usual RWA limit, and we take the resonant case $\omega= \omega_z$, though our approach is not specific to that case. We numerically carry out the (rather computationally demanding) search over the pulse durations $t_1$ and $t_2$, and phases $\Phi_{1}$ and $\Phi_2$.

Displaying the solutions is somewhat difficult, but Fig.~\ref{fig:gate_RWA} gives a visualization of all possible two-pulse solutions for the Hadamard gate in the resonant RWA limit. We show it here as a point of comparison for the strong driving case below.  Furthermore, one can see that the most obvious solution, $\{\Phi_1,\Phi_2, \frac{\omega_x}{2}t_1, \frac{\omega_x}{2}t_2\}=\{\frac{\pi}{2},\pi,\frac{\pi}{2}, \pi\}$, is not the solution with the minimal total time. Rather, there is a solution with $t_1=t_2$ that is the fastest.  The set of solution vectors may be thought of as a closed string in the 4D parameter space. This space is actually a hypertorus, as the RWA Hamiltonian for the two pulses is periodic in both $\Phi_{1,2}$ and $t_{1,2}$.  We can also note in Fig.~\ref{fig:gate_RWA} a symmetry between a pair of solutions at the same $t_{Tot}$: $t_{1,2}\leftrightarrow t_{2,1}$ \& $\Phi_{1,2}\leftrightarrow 2\pi-\Phi_{2,1}$ \cite{footnote_symmetry}. For the $\frac{\pi}{8}$ gate in this RWA limit, the plots are even simpler since the solutions are of the form $\{\Phi_2= \Phi_1 +\frac{\pi}{8}, \frac{\omega_x}{2}t_2= \frac{\omega_x}{2} t_1=\pi \}$. This forms a straight line perpendicular to the $\{t_1, t_2\}$ plane bent in the hypertorus.

\begin{figure*}[!htbp]
\centering
\includegraphics[width=18cm]{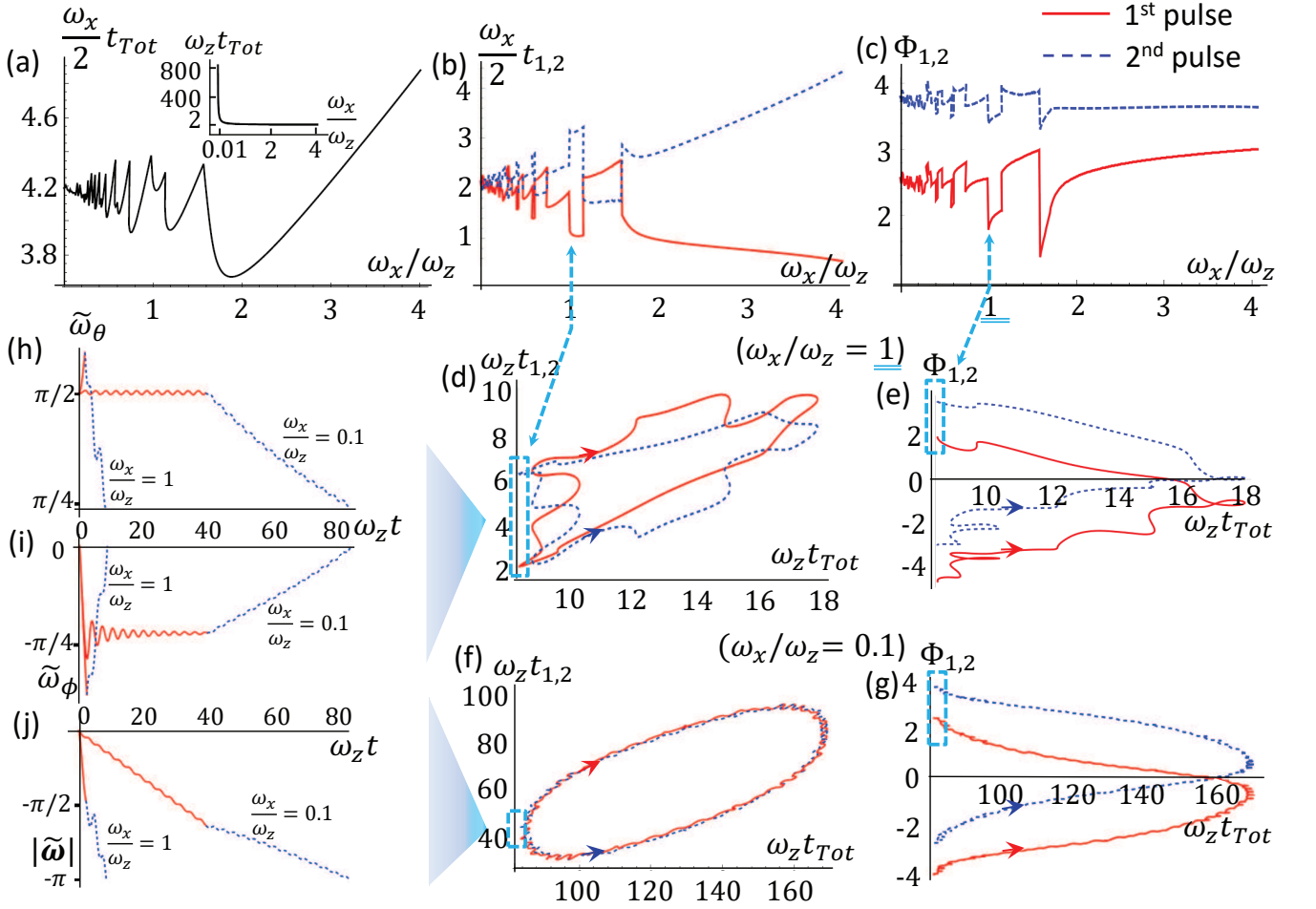}
\caption{Two-pulse solutions for the resonant Hadamard gate with $\omega_x/\omega_z\leq 4$. Times and frequencies are given in units such that $\omega_z=1$. (a) Minimal total time (scaled by driving strength) vs driving strength. The inset shows the same time in terms of the fixed unit. (b) Pulse segment times (scaled by driving strength) vs driving strength. Red solid and blue dashed curves represent the first and second pulses respectively, throughout. (c) Pulse segment phases vs driving strength. (d) Parametric plot of pulse segment times $t_1$ and $t_2$ vs total time for $\omega_x = \omega_z$. (e) Parametric plot of pulse segment phases $\Phi_1$ and $\Phi_2$ vs total time for $\omega_x = \omega_z$. (Note the $2\pi$ periodicity in $\Phi_{1,2}$.) The arrows are a guide to the eyes, indicating which values on the multivalued curves belong to the same solution.  The minimal $t_{Tot}$ solution is enclosed by the dotted cyan box.
(f) Parametric plot of pulse segment times $t_1$ and $t_2$ vs total time for $\omega_x = 0.1\omega_z$. (g) Parametric plot of pulse segment phases $\Phi_1$ and $\Phi_2$ vs total time for $\omega_x = 0.1\omega_z$.
(h)-(j) The evolution operators corresponding to the minimal $t_{Tot}$ solutions for $\omega_x/\omega_z= 0.1$ and 1 vs time. The axial vector $\tilde{\bm{\omega}}$ describes the cumulative effect of the evolution up to time $t$ in terms of an effective axis of rotation, $\{\tilde{\omega}_{\theta}, \tilde{\omega}_{\phi}\}$ and an angle, $|\tilde{\bm{\omega}}|$.
}\label{fig:Sinu_Hada}
\end{figure*}

With this visual representation in mind, we now show the solutions outside the RWA limit. First we consider the Hadamard gate for example. The shape of the string (or strings) of solutions in the 4D variable space now depends on the driving strength, $\omega_x$.   Figures~\ref{fig:Sinu_Hada} (f) and (g) explicitly show the whole solutions for $\omega_x=0.1\omega_z$. As this is fairly weak driving, the solutions are very similar to those in the RWA limit shown in Fig.~\ref{fig:gate_RWA}, with small, rapidly oscillating deviations due to the counter-rotating terms.  The solutions for stronger driving of $\omega_x=\omega_z$ are shown in Figs.~\ref{fig:Sinu_Hada}(d) and \ref{fig:Sinu_Hada}(e). There the distortion compared to the RWA case of Fig.~\ref{fig:gate_RWA} is dramatic.

To give some idea of the dynamics of these highly nontrivial solutions, we use the solution parameters with minimal total gate time and plot their detailed gate evolution in Figs.~\ref{fig:Sinu_Hada}(h)-(j). The evolution matrix after the two pulses, $\tilde{U}_{I\!I}(t_1+t_2)=\tilde{U}_{\rm sin}(t_1)\tilde{U}_{\rm sin}(t_2)$, has to end up equal to $U_H$. We show how this operator evolves in time to the desired one, representing the operator in terms of a rotation axis and angle. In the RWA limit, the evolution operator has a fixed rotation axis and an angle that increases linearly with time in the rotating frame.  Figures~\ref{fig:Sinu_Hada}(h)-(j) depict the total axial vector $\tilde{\bm{\omega}}(t)$'s direction $\{\tilde{\omega}_{\theta}, \tilde{\omega}_{\phi}\}$ and magnitude $|\tilde{\bm{\omega}}|$ during these two pulses, which indeed accomplish $\tilde{\omega}_{\theta}=\pi/4$,  $\tilde{\omega}_{\phi}=0$ and  $|\tilde{\bm{\omega}}|=-\pi$ of $U_H$.  (Note that this indeed occurs much faster for $\omega_x=\omega_z$ than for $\omega_x=0.1\omega_z$.)

\begin{figure*}[!htbp]
\centering
\includegraphics[width=18cm]{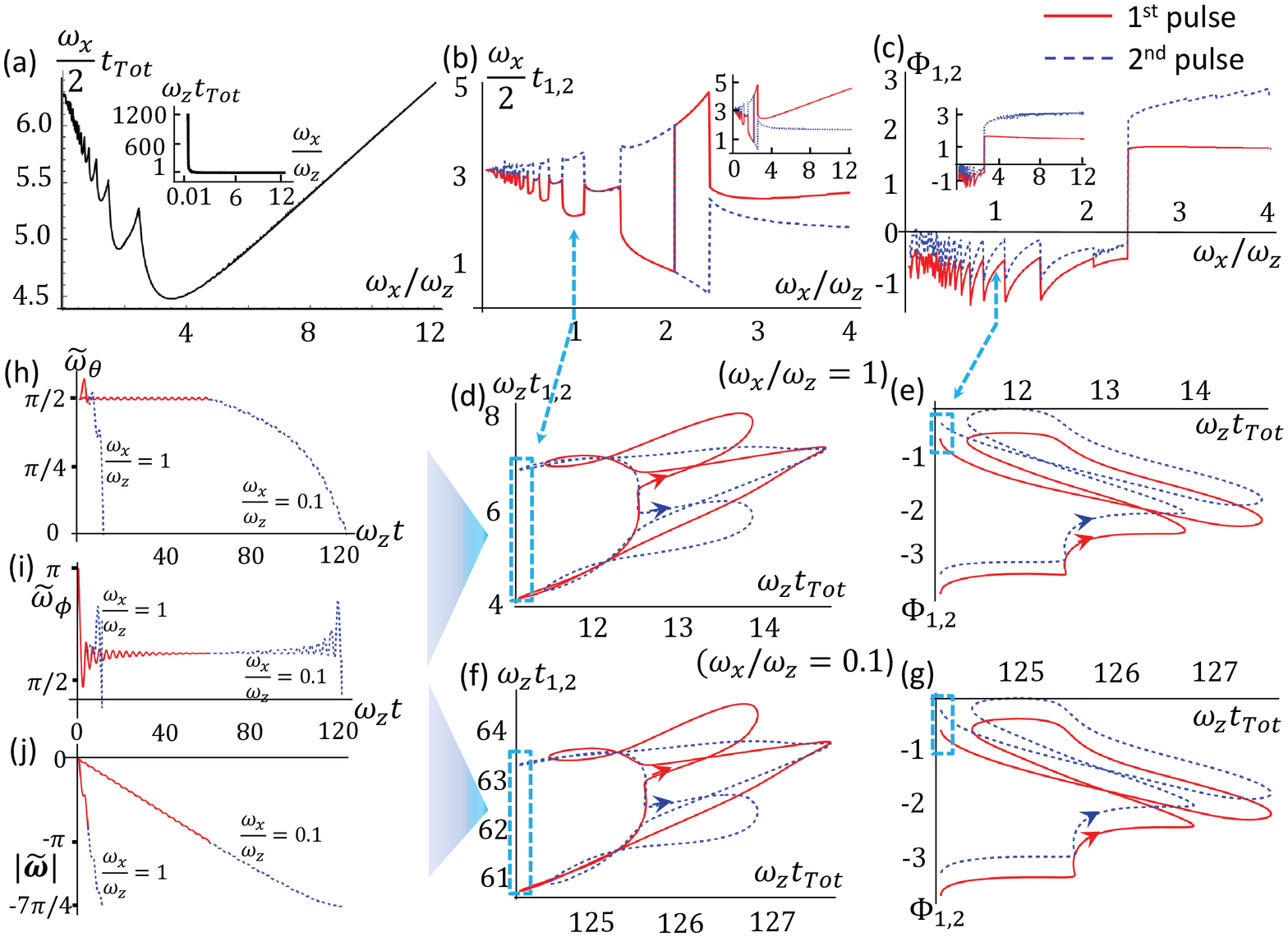}
\caption{Two-pulse solutions for the resonant $\pi/8$ phase gate with $\omega_x/\omega_z\leq 12$. Times and frequencies are given in units such that $\omega_z=1$. (a)-(j) follow the same structure as in Fig.~\ref{fig:Sinu_Hada} for the Hadamard gate. Most features are within $\omega_x\leq4 \omega_z$, but the additional insets in (b) and (c) show the extended range $0<\omega_x\leq12 \omega_z$. (e) and (g) only show $\Phi_{1,2}$ over an interval of $\pi$ as that is the periodicity in this case (see Appendix~\ref{app:shifted_sinu} for details).
}\label{fig:Sinu_PiOver8}
\end{figure*}

Finally, we select the minimal total time solution and map out the parameters of this optimal solution for a wide range of driving amplitude in Figs.~\ref{fig:Sinu_Hada} (b) and (c). We show the minimal total time, scaled by $\omega_x$, as a function of $\omega_x$ in Fig.~\ref{fig:Sinu_Hada} (a). Note that $\omega_x t_{Tot}$ does not vary much as $\omega_x$ increases from zero to the order of $\omega_z$, indicating that the time there is roughly inversely proportional to driving amplitude in that regime.  Around $\omega_x \sim 2\omega_z$, $t_{Tot}$ begins to saturate at around $2\omega_z^{-1}$, as can be seen in the set of Fig.~\ref{fig:Sinu_Hada} (a). As a result, the curve increases linearly, indicating that the time saturates at some constant value regardless of how much more strongly the qubit is driven.  This is not unexpected: In the RWA regime, gate speed scales as $1/\omega_x$, and evidently this scaling roughly holds even up to $\omega_x\sim\omega_z$.  In the other extreme, when $\omega_x\gg \omega_z$, the counter-rotating coupling is nearly stationary within time $1/\omega_x$, and one is in the static coupling limit where the gate speed is limited by $\omega_z$ independent of the value of $\omega_x$.  The transition point between these two regimes evidently depends on the specific logic gate, as we will show below.  The main practical point though is that the duration of a Hadamard gate in the saturated, strongly driven regime is at least two orders of magnitude shorter than in the weakly driven regime where RWA applies.  This should motivate using the fast gate constructions shown in this work. We believe that the optimal fast gate construction outlined here going beyond RWA is the simplest method for achieving FTEC in semiconductor qubit operations.

Figure~\ref{fig:Sinu_PiOver8} presents the same results for the $\frac{\pi}{8}$ phase gate. We again see in Fig.~\ref{fig:Sinu_PiOver8}(a) the similar trend of gate operation speed-up until saturation (as in Fig.~\ref{fig:Sinu_Hada}), though here for the $\pi/8$ gate the saturation occurs at somewhat stronger driving than for the Hadamard gate. The solution string shapes for $\omega_x=0.1\omega_z$ and $\omega_x=\omega_z$ are very similar to each other, although differences become visible when $\omega_x$ further increases (not shown).

We have also found equivalent solutions for an important extension of the standard driving model where the oscillations no longer need be centered at zero, $J_x=\hbar[\omega_{\rm ave}+\omega_x\cos(\omega t+\Phi)]$.
This is relevant for a larger category of qubit systems, and in particular, the ST qubit whose exchange coupling between $|\!\uparrow \downarrow\rangle$ and $|\!\downarrow\uparrow\rangle$ cannot dynamically change sign.  The stationary term $\hbar\omega_{\rm ave}$ is more effective than the counter-rotating part to cause distortion of the Rabi oscillation when $\omega_{\rm ave}$ is not much smaller than $\omega_z$. Notwithstanding, the gate speed-up with increasing $\omega_x$ remains robust.   The detailed results and discussions are given in Appendix~\ref{app:shifted_sinu}. This generic extension also serves to demonstrate the flexibility of our procedure for handling different oscillatory drivings. We can readily modify the coupling model to any experimentally relevant situation and calculate the pulse control parameters for desired logic gates, allowing easy experimental access to fast operations. Our work can also be easily generalized to other qubits where fast operations are desirable (e.g. superconducting qubits).

\section{Conclusion}\label{sec:conclusion}

If desired, semiconductor spin qubits could be operated for faster gate operations in a strong driving regime, far beyond the threshold where RWA gives an accurate description of the dynamics. The strong driving is indeed very desirable to speed up operations, but the design of gate operations in that regime requires appropriate protocols different from the RWA scheme.  This is particularly true given the stringent precision requirements of fault-tolerant quantum computing. In this work we have quantified the well-known limitations of RWA for specific spin qubit platforms by carefully analyzing the experimental situations, and also determined the extent to which a perturbative approach can extend the applicability of an analytic approach to gate design. The analytical approach we developed, which may work when the driving is not too strong (and hence the gate operations are not very fast), is simple and practical to use. We find that for the resonant-exchange and ac-driven singlet-triplet qubits in particular, a completely new approach is required to circumvent the limitations of RWA.

While one could focus on the aspect of minimizing the gate time through any of the standard pulse-shaping or optimal control techniques, in this work we have instead considered the solutions that exist within a simple, highly constrained space.  This has inherent heuristic value, and the ability to map out the full set of solutions is unusual among numerical optimal control approaches, and may prove useful in informing local searches in the experimental parameter space for non-RWA gating pulses. Our desired gate operation is built from only two applications of a simple sinusoidal pulse. Our complete numerical calculations of the pulse control parameters for the Hadamard and $\frac{\pi}{8}$ logic gates have been plotted as a visual look-up table to synthesizing gates over a wide range of driving amplitudes.  While this means of presentation is necessarily imprecise, any real experimental implementation will have imperfections not included in the theoretical model that will perturb the solutions we have presented anyway.  The true power of our results is that they provide a guide to the on-chip calibration of fast gating operations.

For convenience, we provide a quantitative comparison with the ultimate quantum speed limit. It is emphasized that the optimization of quantum gates or evolution operators [on $SU(2)$, or diffeomorphic to the $S^3$ space] \cite{Khaneja_PRA01,  DAlessandro_IEEE01, Wu_PLA02, Boozer_PRA12, Garon_PRA13, Avinadav_PRB14} are different than that of the state transfer which is only parametrized by two real numbers on the Bloch sphere \cite{Boscain_JMP06, Hegerfeldt_PRL13}. In the weak-driving RWA regime, the  minimal times to achieve the Hadamard and $\frac{\pi}{8}$ gates are $3.938/\frac{\Omega}{2}$ and $\frac{\sqrt{15}}{4}\pi/\frac{\Omega}{2}$ ($\hbar\Omega$ is the bound of the control field), respectively, by resorting to Pontryagin maximum principle \cite{Pontryagin_book74,  Garon_PRA13} and minimizing the action \cite{Boozer_PRA12}. The conventional RWA implementation of the Hadamard and $\frac{\pi}{8}$ gates spend longer times, $\frac{3\pi}{2}/\frac{\omega_x}{2}$ and $2\pi/\frac{\omega_x}{2}$, as limited by the fixed control field axis (to be $\hat{\mathbf{x}}$). In the single axis strong-driving regime, the absolute minimal time is $\frac{\pi}{2\omega_z}$ or $\frac{\pi}{4\omega_z}$, for the Hadamard or  $\frac{\pi}{8}$ gate, respectively \cite{Khaneja_PRA01}, whereas our strong-driving $t_{Tot}$ in Figs.~\ref{fig:Sinu_Hada}(a) and \ref{fig:Sinu_PiOver8}(a) are within a factor of 2. Thus, this comparison also serves to demonstrate the overall high performance of our scheme despite of its simple design.

This simple yet highly constrained control pulse is not intended to compete with those using optimal control theory \cite{Krotov_book95, Khaneja_JMR05, Maximov_JCP08}, such as the recent single-qubit implementation by Scheuer \textit{et al} \cite{Scheuer_NJP14}. As a concrete comparison, the method of Ref.~\cite{Scheuer_NJP14} produces high-fidelity state rotations whose speed is less than the relevant quantum speed limit by a mere 3\% by optimizing over 15 parameters and incorporating realistic technical constraints of the pulse generator, while our method produces exact gates whose speed is less than the relevant quantum speed limit by 50\% or less by optimizing over only four parameters in an idealized theoretical scenario.  We hasten to add, however, that in many ways this is an apples-to-oranges comparison, not least because the quantum speed limits for rotations from a specific input state to a specific output state, as in Ref.~\cite{Scheuer_NJP14}, are different from the speed limits for implementing a state-independent gate operation (in fact, the latter are not even fully worked out yet \cite{Avinadav_PRB14}).  Nonetheless, this illustrates the point that for a specific goal or cost functional (i.e., some combination involving fidelity, time, bandwidth, etc.), techniques such as that of Scheuer \textit{et al} \cite{Scheuer_NJP14} may offer better performance.  However, the attraction of the pulses we have presented is their theoretically rudimentary form and the ability to map out the entire solution set within the comparatively small parameter space in order to aid calibration.

As environmental dephasing errors increase with operation time, the two orders of magnitude speed-up permitted by non-RWA pulses is a promising tool for enabling fault-tolerant quantum computation with spin qubits. For a given experiment with specified constraints and objectives, a more sophisticated approach may even allow further speed-up or incorporate additional robustness. Going beyond RWA will eventually be essential for resonant exchange and singlet-triplet qubits in order to reach the error correction thresholds for quantum computation.

This work is supported by LPS-MPO-CMTC.  XW acknowledges support from City University of Hong Kong (Projects No. 9610335 and No. 7200456).

\appendix
\section{LOGIC GATES REALIZATION WITH SHIFTED-SINUSOIDAL COUPLING}\label{app:shifted_sinu}
\begin{figure*}[!hbtp]
\centering
\includegraphics[width=18cm]{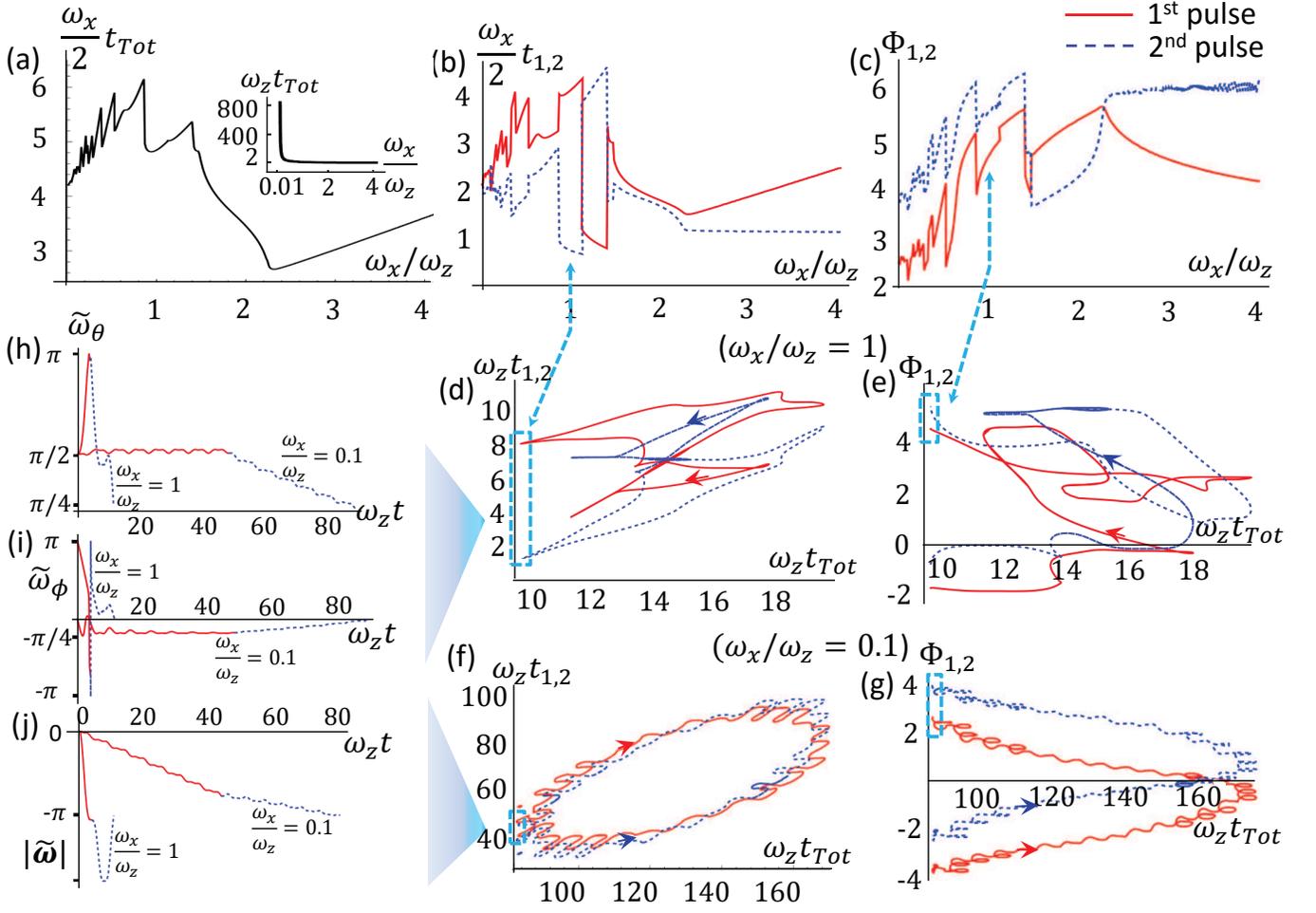}
\caption{Two-pulse solutions for the resonant Hadamard gate with $\omega_x/\omega_z\leq 4$ for the shifted oscillations of Eq.~\eqref{eq:shifted_Jx} with $\omega_{\rm ave}=2\omega_x$. Times and frequencies are given in units such that $\omega_z=1$. The arrangement of the Figs. (a)-(j) and notations follow those in Fig.~\ref{fig:Sinu_Hada}. In (i), the abrupt change of $\tilde{\omega}_\phi$ in the $\omega_x=1 \omega_z$ case, i.e., the jump from $-\pi$ to $\pi$ at around $t=4$, is simply due to the choice $-\pi<\tilde{\omega}_\phi\leq \pi$ here.
}\label{fig:Hadamard_gate_shift_sinu}
\end{figure*}
\begin{figure*}[!htbp]
\centering
\includegraphics[width=18cm]{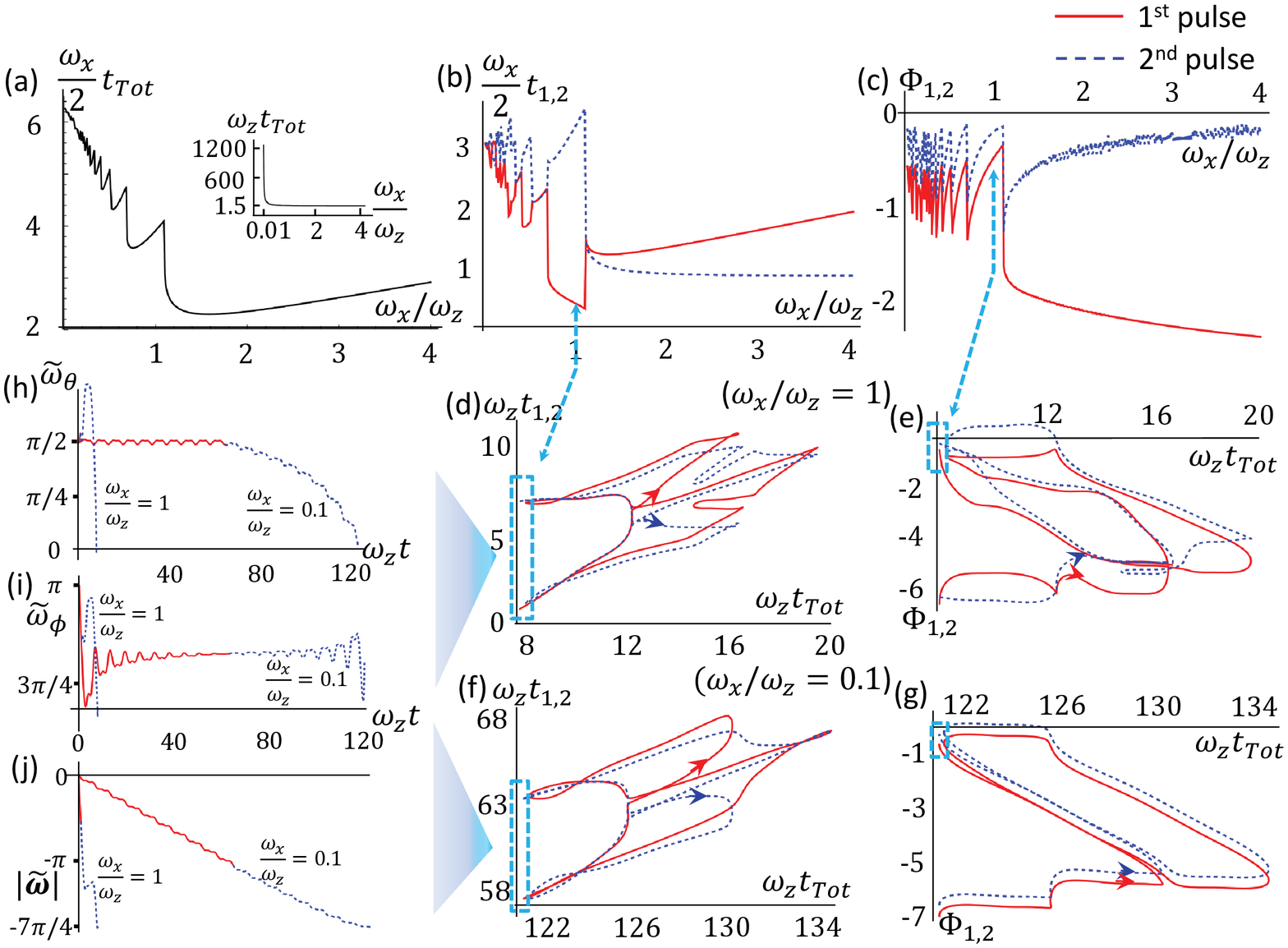}
\caption{Two-pulse solutions for the resonant $\pi/8$ phase gate with $\omega_x/\omega_z\leq 4$ for the shifted oscillations  of Eq.~\eqref{eq:shifted_Jx} with $\omega_{\rm ave}=2\omega_x$. Times and frequencies are given in units such that $\omega_z=1$. The arrangement of the Figs. (a)-(j) and notations follow those in Fig.~\ref{fig:Sinu_Hada}.
}\label{fig:PiOver8_gate_shift_sinu}
\end{figure*}

In this appendix, we present the realization of the Hadamard and $\frac{\pi}{8}$ gates by shifted sinusoidal coupling,
\begin{equation}\label{eq:shifted_Jx}
J_x=\hbar\left[\omega_{\rm ave}+\omega_x\cos(\omega t+\Phi)\right].
\end{equation}
Considering this experimentally relevant and generic modification also demonstrates that our procedure for providing pulse control knobs for quantum logic gates can apply to more complicated coupling functions in the Hamiltonian, and that the general trend of gate speed-up is not specific to the case considered in the main text.

Figures~\ref{fig:Hadamard_gate_shift_sinu} and \ref{fig:PiOver8_gate_shift_sinu} are specific results for the Hadamard and $\frac{\pi}{8}$ gates, respectively, at $\omega_{\rm ave}=2\omega_x$. We arrange them in the same way as in Figs.~\ref{fig:Sinu_Hada} and \ref{fig:Sinu_PiOver8} for convenient comparison. In the limit of small $J_x/J_z$, intuitively the important term in driving the qubit is again the in-phase (RWA) term. Both the counter-rotating and the stationary shift terms are suppressed following the spirit of RWA. In the rotating frame, where the in-phase part provides a fixed rotation axis, the $\hbar\omega_{\rm ave}$ term rotates at angular frequency $\omega$, and thus becomes ineffective in precessing the qubit when $\omega_{\rm ave}\ll \omega=\omega_z$ (the latter is set to be the unit frequency in the figures). This conclusion also applies for any low frequency (or far off-resonant) noise when its magnitude is small compared to the two-level splitting.

Focusing on the Hadamard gate, when $\omega_x$ grows to 0.1 $\omega_z$, as Figs.~\ref{fig:Hadamard_gate_shift_sinu} (f) and (g) show, the distortion on the RWA solution (see Fig.~\ref{fig:gate_RWA}) is appreciable, while the main shape is largely preserved. The main frequency of the oscillatory distortion emerges as half of that in the unshifted of Figs.~\ref{fig:Sinu_Hada} (f) and (g). This is because the stationary shift ($\hbar\omega_{\rm ave}$) is more effective than the counter-rotating part ($\hbar \omega_x e^{i\omega t}$) because (I) the former rotates slower ($\omega_z$) than the latter ($2\omega_z$) in the rotating frame and (II) $\omega_{\rm ave}>\omega_x$ in our model ($\omega_{\rm ave}/\omega_x=2$). At $\omega_x=1$ [Figs.~\ref{fig:Hadamard_gate_shift_sinu} (d) and (e)], the distortion by $\hbar\omega_{\rm ave}$ becomes evidently overwhelming compared to the unshifted case of Figs.~\ref{fig:Sinu_Hada} (d) and (e), and gradually chaotic at some locations. For most parts on the strings formed by the solution vector ($\{\Phi_1,\Phi_2, t_1, t_2\}$), though, the solution remain convergent towards these strings in phase space.

The results for the $\frac{\pi}{8}$ gate are similar. Comparing Fig.~\ref{fig:PiOver8_gate_shift_sinu} (a) with Fig.~\ref{fig:Sinu_PiOver8} (a), one notable aspect is more rapid shortening of the gate time in the shifted sinusoidal case.

A more interesting difference with the unshifted case is the doubling of the solution vector periodicity along $\Phi_{1,2}$ directions.  In fact, it is a peculiar feature of the phase gate for purely sinusoidal coupling that the periodicity for the solution string is $\pi$ along the $\Phi_1$ and $\Phi_2$ dimensions [see Figs.~\ref{fig:Sinu_PiOver8} (e) and (g)], rather than the trivial $2\pi$.  That stems from its reversing upon $\Phi_{1,2}\rightarrow\Phi_{1,2}+\pi$.  This feature is explained by the commutativity of $\pi_{z}$ (where $\pi_{z}=[-i, 0 \text{;} 0, i]$ is a $\pi$ rotation about $z$) and the phase gate operation, $\pi_{z}^\dag U_T \pi_{z}=U_T$. The net operation on the qubit can be described by an  axial vector $\tilde{\bm{\omega}}\|\mathbf{z}$; a shift of  $\Phi_{1,2}$ by $\pi$ reverses the driving term since $\cos(\omega t+\Phi+\pi)=-\cos(\omega t+\Phi)$ and amounts to an additional $\pi$ rotation of the driving field around the $z$ axis in the rotating frame; obviously this $\pi$ rotation has no effect on  $\tilde{\bm{\omega}}$ and hence the ensuring periodicity for $\Phi_{1,2}$. It applies for phase shift gate of arbitrary angles. The addition stationary term in $J_x$ breaks this relation and thus returns the periodicity with $\Phi_{1,2}$ to the normal $2\pi$.

Finally, we note that the shifting term also breaks the shape resemblance of the solution string between $\omega_x=0.1$ and 1 $\omega_z$ cases that exist in Fig.~\ref{fig:Sinu_PiOver8}.


\begin{thebibliography}{99}
\bibitem{footnote_symmetry} It is unique for the Hadamard gate under RWA: these two solutions are related by $\pi_{y}$ ($\pi$ rotation around the $y$ axis), as $(\pi_{y}U_H\pi_y^\dag)^\dag = U_H$.
\end{thebibliography}
\end{document}